\documentstyle[12pt]{article}

\begin{document} 
\vskip 2cm
\begin{center}
{\large Momentum and doping dependence of the electron
spectrum and ($\pi$,0) feature in copper oxide materials}\\
\bigskip
Feng Yuan$^{a}$, Xianglin Ke$^{a}$, and Shiping Feng$^{a,b}$\\
\bigskip
$^{a}$Department of Physics, Beijing Normal University, Beijing 100875,
China\\
\bigskip
$^{b}$National Laboratory of Superconductivity, Academia
Sinica, Beijing 100080, China\\
\bigskip
\end{center}

The momentum and doping dependence of the electron spectrum of
copper oxide materials in the underdoped regime is studied
within the $t$-$J$ model. It is shown that the pseudogap opens
near ($\pi$,0) point in the Brillouin zone, and the electron
dispersion exhibits the flat band around ($\pi$,0), which are
consistent with the experiments.
\bigskip

The experimental measurements from the photoemission spectroscopy
[1] show that electron spectrum in copper oxide materials is
strongly momentum and doping dependent, and has an anomalous form
as a function of energy $\omega$ for ${\bf k}$ in the vicinity of
(${\pi}$,0) point in the Brillouin zone, which leads to the flat
band near momentum ($\pi$,0) with anomalously small changes of
electron energy as a function of momentum. This flat band around
($\pi$,0) point has a particular importance in the mechanism of
the normal-state pseudogap formation [1]. Although the exact origin
of the flat band still is controversial, a strongly correlated
many-body like approach may be appropriate to describe the electronic
structure of copper oxide materials. In this paper, we study this
issue within the $t$-$J$ model [2],
\begin{eqnarray}
H=-t\sum_{i\hat{\eta}\sigma}C^{\dagger}_{i\sigma}C_{i+\hat{\eta}
\sigma}+J\sum_{i\hat{\eta}}{\bf S}_{i}\cdot {\bf S}_{i+\hat{\eta}},
\end{eqnarray}
acting on the space with no doubly occupied sites,
where $\hat{\eta}=\pm \hat{x},\pm\hat{y}$, $C^{\dagger}_{i\sigma}$
($C_{i\sigma}$) are the electron creation (annihilation) operators,
and ${\bf S}_{i}$ is spin operator. The strong electron correlation
in the $t$-$J$ model manifests itself by the electron single
occupancy on-site local constraint [2]. To incorporate this local
constraint, the fermion-spin theory based on the charge-spin
separation, $C_{i\uparrow}=h^{\dagger}_{i}S^{-}_{i}$,
$C_{i\downarrow}=h^{\dagger}_{i}S^{+}_{i}$, has been proposed [3],
where the spinless fermion operator $h_{i}$ keeps track of the charge
(holon), while the pseudospin operator $S_{i}$ keeps track of the
spin (spinon). In this paper, we hope to discuss the electronic
structure of copper oxide materials, and therefore it needs to
calculate the electron Green's function $G(i-j,t-t')=\langle\langle
C_{i\sigma}(t);C^{\dagger}_{j\sigma}(t')\rangle\rangle$. According to
the fermion-spin theory, the electron Green's function is a
convolution of the spinon Green's function
$D(i-j,t-t')=\langle\langle S^{+}_{i}(t);S^{-}_{j}(t')\rangle\rangle$
and holon Green's function $g(i-j,t-t')=\langle\langle h_{i}(t);
h^{\dagger}_{j}(t')\rangle\rangle$, and can be formally expressed
in terms of the spectral representation as,
\begin{eqnarray}
G({\bf k},\omega)={1\over N}\sum_{q}\int^{\infty}_{-\infty}
{d\omega' \over 2\pi}\int^{\infty}_{-\infty}{d\omega'' \over 2\pi}
A_{h}({\bf q}, \omega')\times\nonumber \\
A_{s}({\bf q+k},\omega''){n_{F}(\omega')+
n_{B}(\omega'')\over \omega+\omega'-\omega''},~~~~(2)
\end{eqnarray}
where the holon spectral function
$A_{h}({\bf q},\omega)=-2{\rm Im}g({\bf q},\omega)$, spinon spectral
function $A_{s}({\bf k},\omega)=-$ $2{\rm Im}D({\bf k},\omega)$,
and $n_{B}(\omega)$ and $n_{F}(\omega)$ are the boson and fermion
distribution functions for spinons and holons, respectively. In this
paper, we limit the spinon part to the first-order (mean-field level)
since some physical properties can be well described at this level
[4]. On the other hand, it has been shown that there is a connection
between the anomalously temperature dependence of the resistivity and
the flat band around ($\pi$,0) point and normal-state pseudogap [1].
Therefore we treat the holon part by using the loop expansion to the
second-order correction as in the discussion of the charge dynamics
[4]. Within the fermion-spin theory, the mean-field spinon and full
holon Green's functions $D^{(0)}({\bf k}, \omega)$ and $g({\bf k},
\omega)$ have been evaluated in Ref. [5] and Ref. [4], respectively.
Substituting these Green's functions into Eq. (2), we therefore can
obtain the electron spectral function
$A({\bf k},\omega)=-2{\rm Im}G({\bf k},\omega)$.

The numerical results of the electron spectral function at ($\pi$,0)
point in the doping $\delta =0.06$ for the parameter $t/J=2.5$ in the
zero temperature are plotted in Fig. 1 (solid line). For comparison,
the corresponding mean-field results [5] (dashed line) are also
plotted in Fig. 1. These results indicate that at the mean-field level,
the electron spectrum at ($\pi$,0) point consists of two main
parts, which comes from noninteracting particles. After including the
fluctuation, the mean-field part is renormalized and the spectral
weight has been spread to lower energies, in particular, the sharp
mean-field peak at ($\pi$,0) point near the chemical potential $\mu$
has been split into two peaks. Moreover, the low energy peaks are well
defined at all momenta, and the positions of the dominant peaks in
$A({\bf k},\omega)$ as a function of momentum in the doping $\delta=
0.06$ for the parameter $t/J=2.5$ are shown in Fig. 2 (solid line).
In comparison with corresponding mean-field results [5] (dashed line)
in Fig. 2, it is shown that the mean-field electron dispersion
$E^{(0)}_{k}$ in the vicinity of ($\pi$,0) point has been split into
two branches $E^{(-)}_{k}$ and $E^{(+)}_{k}$, and a pseudogap
opens. The branch $E^{(-)}_{k}$ has a very weak dispersion around
($\pi$,0) point, and then the flat regime appears, while the Fermi
energy is only slightly above this flat regime. Although the nature
of the pseudogap is different in different theories, the present
results show that the pseudogap near ($\pi$,0) point is closely
related to the spinon fluctuation, since the full holon Green's
function (then the electron Green's function) is obtained by
considering the second-order correction due to the spinon pair bubble,
where the single-particle hopping is strongly renormalized by the
short-range antiferromagnetic order resulting in a bandwidth also of
order of (a few) $J$, this renormalization is then responsible for
the anomalous dispersion around ($\pi$,0) point and normal-state
pseudogap.

\vskip 2cm
\centerline{Acknowledgements}

The authors would like to thank Professor C.D. Gong, Professor H.Q.
Lin, and Professor Z.X. Zhao for helpful discussions. This work was
supported by the National Natural Science Foundation of China under
Grant No. 19774014.

\newpage
\begin{enumerate}
\item Z.X. Shen {\it et al.}, Phys. Rep. 253 (1995) 1.
\item P.W. Anderson, Science 235 (1987) 1196.
\item S. Feng {\it et al.}, Phys. Rev. B49 (1994) 2368.
\item S. Feng {\it et al.}, Phys. Lett. A232 (1997) 293.
\item S. Feng {\it et al.}, Phys. Rev. B55 (1997) 642.

\end{enumerate}

\newpage
\centerline{Figure Captions}

Figure 1. $A({\bf k},\omega )$ at ($\pi$,0) point at the doping
$\delta =0.06$ for $t/J=2.5$. The dashed line is the result at the
mean-field level.

Figure 2. Position of the dominant peaks in $A({\bf k},\omega)$
as a function of momentum at the doping $\delta=0.06$ for $t/J=2.5$.
The dashed line is the result at the mean-field level.

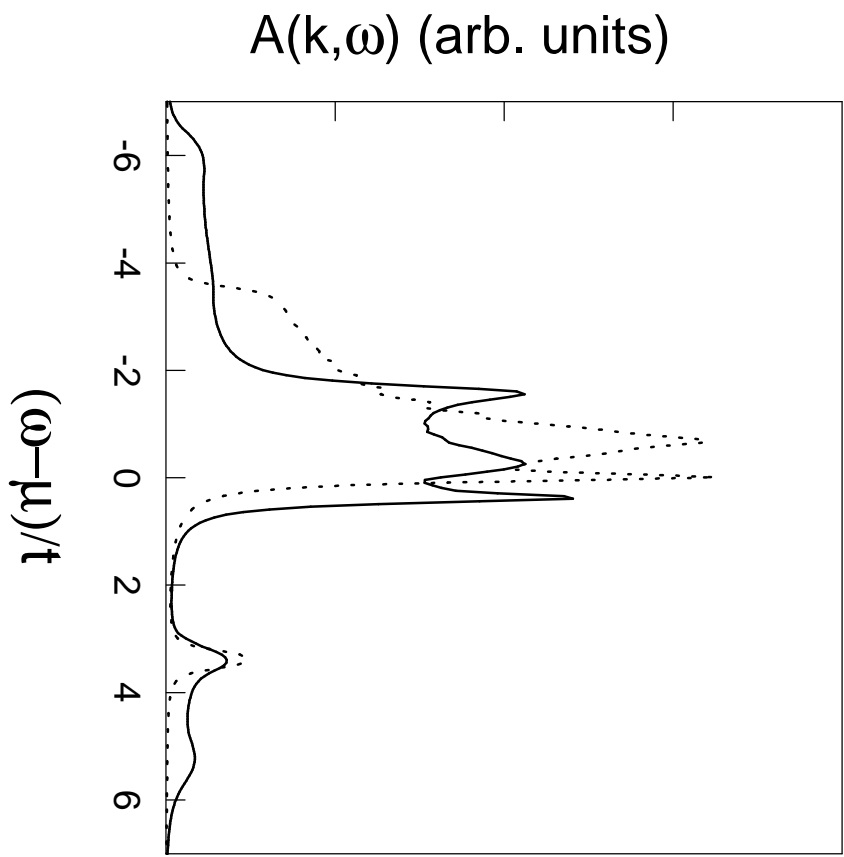
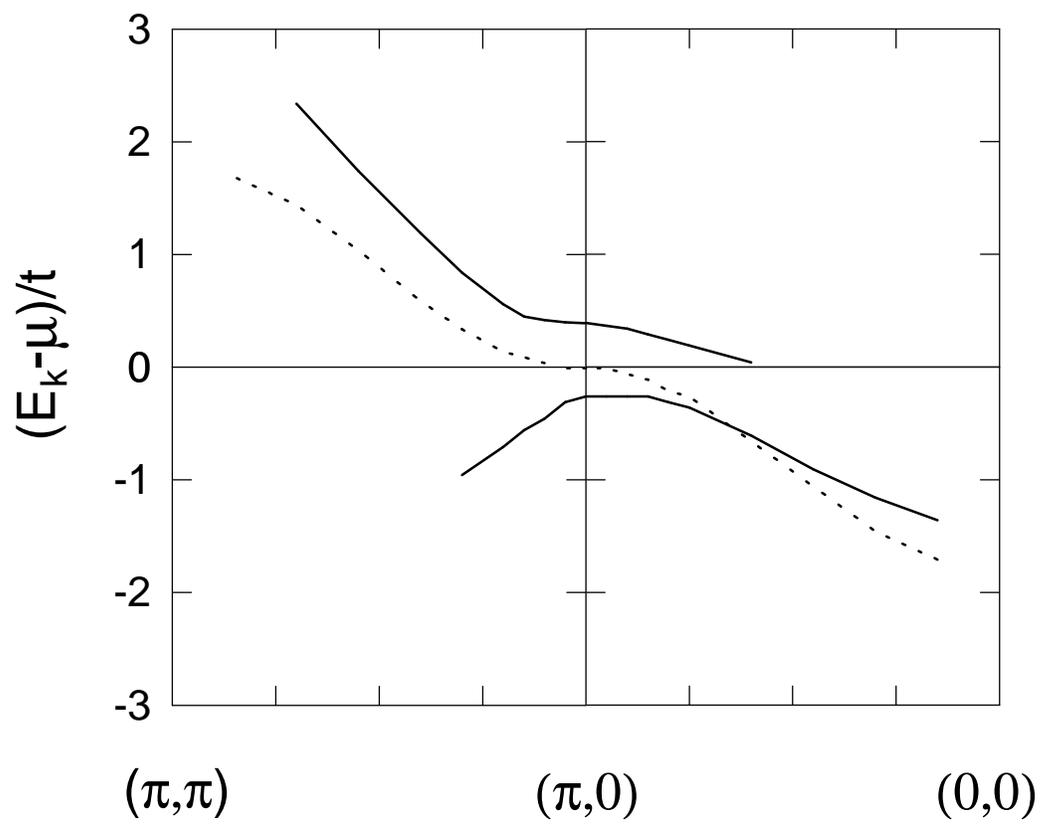
\end{document}